\documentclass[preprint,preprintnumbers,amsmath,amssymb]{revtex4}

\usepackage{dcolumn}
\usepackage{bm}
\ifx\pdfoutput\undefined
   \usepackage[dvips]{graphicx}
\else
   \usepackage[pdftex]{graphicx}
   \DeclareGraphicsExtensions{.jpg,.pdf,.mps,.png}
\fi

\begin{document}

\title{Intrinsic ripples in graphene}

\author{A. Fasolino, J. H. Los and M. I. Katsnelson}

\address {Institute for Molecules and Materials, Radboud
University Nijmegen, 6525 ED Nijmegen, The Netherlands}
\date{\today}


\maketitle

{\bf The stability of two-dimensional (2D) layers and membranes is
subject of a long standing theoretical debate. According to the so
called Mermin-Wagner theorem~\cite{mermin}, long wavelength
fluctuations destroy the long-range order for 2D crystals.
Similarly, 2D membranes embedded in a 3D space have a tendency
to be crumpled~\cite{nelson}. These dangerous fluctuations can,
however, be suppressed by anharmonic coupling between bending and
stretching modes making that a two-dimensional membrane can exist
but should present strong height
fluctuations~\cite{nelson,peliti,radz}. The discovery of graphene,
the first truly 2D crystal~\cite{kostya0,kostya1} and the recent
experimental observation of ripples in freely hanging
graphene~\cite{jannik} makes these issues especially important.
Beside the academic interest, understanding the mechanisms of
stability of graphene is crucial for understanding electronic
transport in this material that is attracting so much interest for
its unusual Dirac spectrum and electronic
properties~\cite{kostya2,kim,reviewGK,reviewktsn}. Here we address
the nature of these height fluctuations by means of
straightforward atomistic Monte Carlo simulations based on a very
accurate  many-body interatomic potential for carbon~\cite{los1}.
We find that ripples spontaneously appear due to thermal
fluctuations with a size distribution peaked around 70\AA~ which
is compatible with experimental findings~\cite{jannik} (50-100 \AA) but not with 
the current understanding of stability of flexible 
membranes~\cite{nelson,peliti,radz}. 
This unexpected result seems to be due to the multiplicity of chemical bonding 
in carbon. }

The phenomenological theories for flexible membranes~\cite{nelson,peliti,radz}
are derived in the continuum limit without including any microscopic feature and
their applicability to graphene in the interesting range of
temperatures, sample sizes etc. is not evident. We present Monte
Carlo simulations of the equilibrium structure of single layer
graphene. By monitoring the normal-normal correlation functions we can
directly compare our results to the predictions of the existing theories.
The effect of interest is crucially dependent on
acoustical phonons and interactions between them and therefore the
simulations require samples much larger than interatomic
distances, typically many  thousands of atoms at thermal
equilibrium, making prohibitive ab-initio simulations a la
Car-Parrinello~\cite{CP}. However, for carbon a very accurate
description of energetic and thermodynamic properties of different
phases is provided  by the effective many body potential
LCBOPII~\cite{los1,los2}. This bond order  potential is constructed in such a
way as to provide a unified description of the energetics and elastic constants of all carbon
phases as well as the energy characteristics  of
different defects with accuracy comparable to experimental
accuracy. We find clear deviations from harmonic behavior for long
wavelength fluctuations but, instead of the
expected power-law scaling at long wavelengths, we find a marked
maximum of fluctuations with wavelength of about $70$ \AA. Against the expectations, 
we also find a stiffening of the bending rigidity with increasing temperature.  
We relate these features to  fluctuations in bond length that in carbon 
signal a partial change from conjugated to single/double bonds, with consequent 
deviations from planarity. 

We perform atomistic
Monte Carlo (MC) simulations based on the standard Metropolis algorithm
for approximately squared samples of different sizes (see Table I)
with periodic boundary conditions. We always start with completely
flat graphene layers to avoid any bias. Typically we used 400000 
MC steps (1 MC step corresponds to $N$ attempts to a coordinate change) to equilibrate
and half a million for averaging. Every 5 MC steps,  we allow for isotropic
size fluctuations. The energy of a given configuration is evaluated
according to the bond order potential LCBOPII developed in the
recent past~\cite{los1}. This potential is based on a large
database of experimental and theoretical data for molecules and solids 
and has been proven to describe very well thermodynamic and structural properties of
all phases of carbon and its phase diagram in a wide range of
temperatures and pressures~\cite{los1,los2}. We believe
that LCBOPII can give a good description of graphene because it
reproduces correctly the elastic properties of graphene and yields
structure and energetics of vacancies in graphene in good
agreement with {\it ab initio} calculations~\cite{carlsson}. Of particular importance here, 
is that  bond order potentials correlate coordination, bond length and bond strength, 
allowing changes between single, double and conjugated bonds with the correct energetics. 
Most
simulations have been performed at room temperature $T=300$ K but
we have also simulated some of the structures at very high
temperature $T=3500$ K close to the bulk graphite melting to study
qualitatively temperature effects.

\begin{figure}
\includegraphics[angle=90,width=18cm]{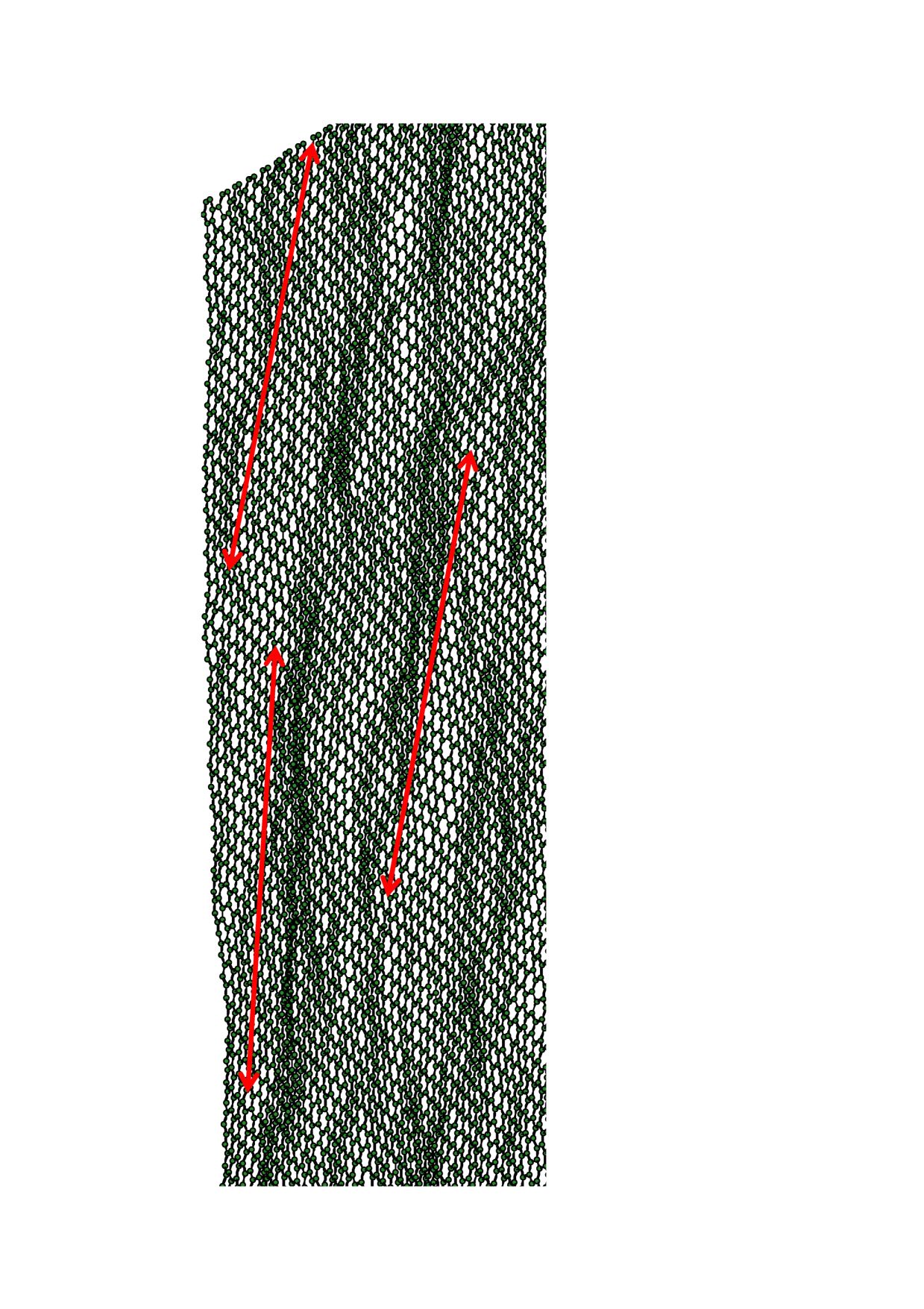}
\caption{Snapshot of the $N=8640$ sample at $T=300$ K. The red arrows are 70 \AA~ long. }
\label{snap}
\end{figure}

A typical snapshot of graphene at room temperature is shown in
Fig.\ref{snap}. The first thing to notice is that  height fluctuations are
present at equilibrium. We find a very broad distribution of
height displacements $\overline{h}$ with a typical size of order of
$0.6$~\AA~ for the $N=8640$ sample, comparable to the interatomic
distance $1.42$ \AA. 

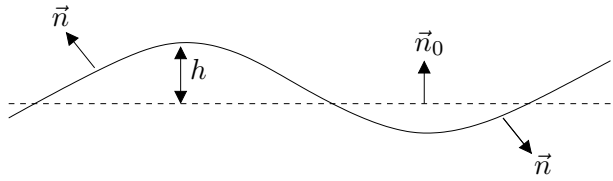
\begin{figure}
\begin{picture}(0,0)%
\includegraphics{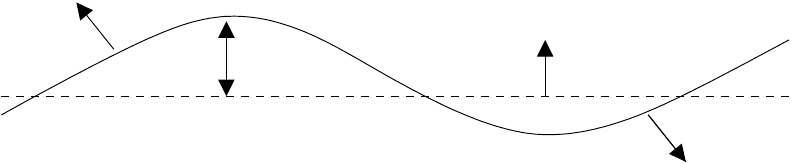}%
\end{picture}%
\setlength{\unitlength}{2368sp}%
\begingroup\makeatletter\ifx\SetFigFont\undefined%
\gdef\SetFigFont#1#2#3#4#5{%
  \reset@font\fontsize{#1}{#2pt}%
  \fontfamily{#3}\fontseries{#4}\fontshape{#5}%
  \selectfont}%
\fi\endgroup%
\begin{picture}(6324,1299)(1789,-5398)
\put(2250,-4050){$\vec{n}$}
\put(3700,-4600){$h$}
\put(6050,-4310){$\vec{n}_0$}
\put(7300,-5600){$\vec{n}$}
\end{picture}%
\caption{Sketch of a flexible membrane (solid line). $h$ is the
out of plane deviation with respect to the $z=0$-plane (dashed
line) defined by the center of mass. The unit vector  $\vec{n}$
and $\vec{n}_0$ are the normals to each point in the membrane and
in the reference plane respectively.}
\label{sketch}
\end{figure}

The natural way to analyse further our results is to compare them to the results and predictions of the phenomenological theories of thermal fluctuations in flexible 
membranes~\cite{nelson,peliti,radz}. To this purpose it is worth to review in some detail 
the main ideas and results of these theories~\cite{nelson,peliti,radz} that are not of common knowledge
outside the soft condensed matter community. The primary
quantities are the two-component displacement vector in the plane
${\bf u}$, the out of plane displacement out of plane $h$ and the
normal unit vector ${\bf n}$ with in-plane component $-\nabla h/
\sqrt{1+ \left(\nabla h\right)^2}$ illustrated in Fig.\ref{sketch}. The
elastic energy of the membrane is given by
\begin{equation}
E=\int d^2x\left[ \frac \kappa 2\left( \nabla ^2h\right) ^2+\mu
\overline{u}_{\alpha \beta }^2+\frac \lambda 2\overline{u}_{\alpha \alpha
}^2\right]
\end{equation}
where $\kappa$ is the bending rigidity, $\mu$ and $\lambda$ are Lam\'{e}
parameters and  $\overline{u}_{\alpha \beta }$ is the deformation tensor:
\begin{equation}
\overline{u}_{\alpha \beta }=\frac 12\left( \frac{\partial u_\alpha }{%
\partial x_\beta }+\frac{\partial u_\beta }{\partial x_\alpha }+\frac{%
\partial h}{\partial x_\alpha }\frac{\partial h}{\partial x_\beta }\right) .
\label{deftens}
\end{equation}

In harmonic approximation, by neglecting the last, non linear,
term in the deformation tensor (\ref{deftens}), the bending ($h$)
and stretching (${\bf u}$) modes are decoupled. In this
approximation the Fourier components of the bending correlation
function with wavevector ${\bf q}$ is
\begin{equation}
\left\langle \left| h_{\mathbf{q}}\right| ^2\right\rangle =\frac {T N}{\kappa S_0 q^4}
\end{equation}
where $N$ is the number of atoms,  $S_0=L_x L_y/N$ is the area per atom
and $T$ is the temperature in units of energy. In this
approximation, the mean square displacement in the direction
normal to the layer is
\begin{equation}
\left\langle  h^2 \right\rangle= \sum_{{\bf q}}\left\langle \left| h_{\mathbf{q}}\right| ^2\right\rangle \propto
\frac {T }{\kappa} L^2
\label{h2harm}
\end{equation}
where $L$ is a typical linear sample size. The correlation function of
the normals
\begin{equation}
G\left( q\right) =\left\langle \left| \mathbf{n}_{\mathbf{q}}\right|
^2\right\rangle =q^2\left\langle \left| h_{\mathbf{q}}\right|
^2\right\rangle
\end{equation}
in this approximation becomes
\begin{equation}
G_0\left( q\right) =\frac {T N}{\kappa S_0 q^2}
\label{G0}
\end{equation}
which implies that the mean square angle between the normals is
logarithmically divergent as $L \rightarrow \infty$~\cite{nelson}. This
behaviour indicates the tendency to crumpling of membranes
due to thermal fluctuations.

Deviations from this harmonic behaviour, namely anharmonic coupling
between bending and stretching modes, can stabilize the flat phase by 
suppressing the long wavelength fluctuations~\cite{nelson,peliti,radz}.
In this case the corresponding correlation
function of the normals is given by the Dyson equation
\begin{equation}
G_a^{-1}\left( q\right) =G_0^{-1}\left( q\right) +\Sigma \left( q\right)
\label{Gq}
\end{equation}
with self energy
\begin{equation}
\Sigma \left( q\right) =\frac {A S_0}{N q^2}\left( \frac{q}{q_0}\right) ^\eta
\label{Sigma}
\end{equation}
where $q_0=2\pi \sqrt{B/\kappa }$, $B$ being the two-dimensional
bulk modulus, $\eta$ is the anomalous rigidity exponent and $A$ is
a numerical factor.

The simplest way to derive this expression is to use the
self-consistent perturbation theory~\cite{radz} which gives
$\eta\approx 0.8$ in reasonable agreement with the
results of Monte Carlo simulation for a model of tethered
membranes~\cite{bowick}. However, Eqs.(\ref{Gq}),(\ref{Sigma})
can be written from a more
general scaling consideration~\cite{peliti}, $q_0$ being the only
factor with dimension of an inverse length that can be constructed
from relevant parameters of this theory. As a result of this
anharmonic coupling, the typical height of fluctuations in the
direction normal to the membrane is much smaller than the one
given by Eq.(\ref{h2harm}) and scales with the sample size as
$L^{\zeta}$, with $\zeta= 1 - \eta/2$. Nevertheless, the
fluctuations are still anomalously large and they can be much
larger than the interatomic distance for large samples. Thus, the
theory predicts an intrinsic tendency to ripple formation. At the
same time, the amplitude $\overline{h}\propto L^\zeta$ of these
transverse fluctuations  remains much smaller than the sample size
and preserves the long-range order of the normals so that the
membrane can be considered as approximately flat and not crumpled.

Another structural issue is the existence of long range
crystallographic order in membranes that can be destroyed by a
finite concentration of topological defects, namely dislocations
and disclinations. For the case of 2D crystals in 2D space both
types of defects have infinite energy as $L\rightarrow \infty$:
the elastic energy of dislocations grows as $\ln(L)$, and of
disclinations as $L$. A general analysis~\cite{nelson} (Chap. 6)
for flexible membranes shows that these divergencies are
suppressed by bending in such a way that the energy of
disclinations behaves as $\ln(L)$ whereas that of  dislocations
remains finite, and of the order of $\kappa$. This means that the
orientational order survives whereas translational order is
destroyed by spontaneous creation of dislocations. However, the
corresponding correlation length is $\propto \exp(const\cdot
\kappa / T)$ which makes this mechanism completely irrelevant for
covalently bonded layers such as graphene, with $\kappa\approx
1.17$ eV~\cite{nicklow}. Indeed,  we never observe any topological
defect in our simulations also at very high temperature,  nor any
experimental evidence of their existence has been
reported~\cite{jannik}.

To obtain  a quantitative comparison with these theoretical predictions for the
spatial distribution of the ripples, we have calculated
numerically the Fourier components of the correlation function of
the normals $G(q)$ for $q_x$ and $q_y$ multiples of $2 \pi/L_x$ and
$2 \pi/L_y$ respectively. To our surprise, our numerical  results
are not described by this general theory.

\begin{figure}
\includegraphics{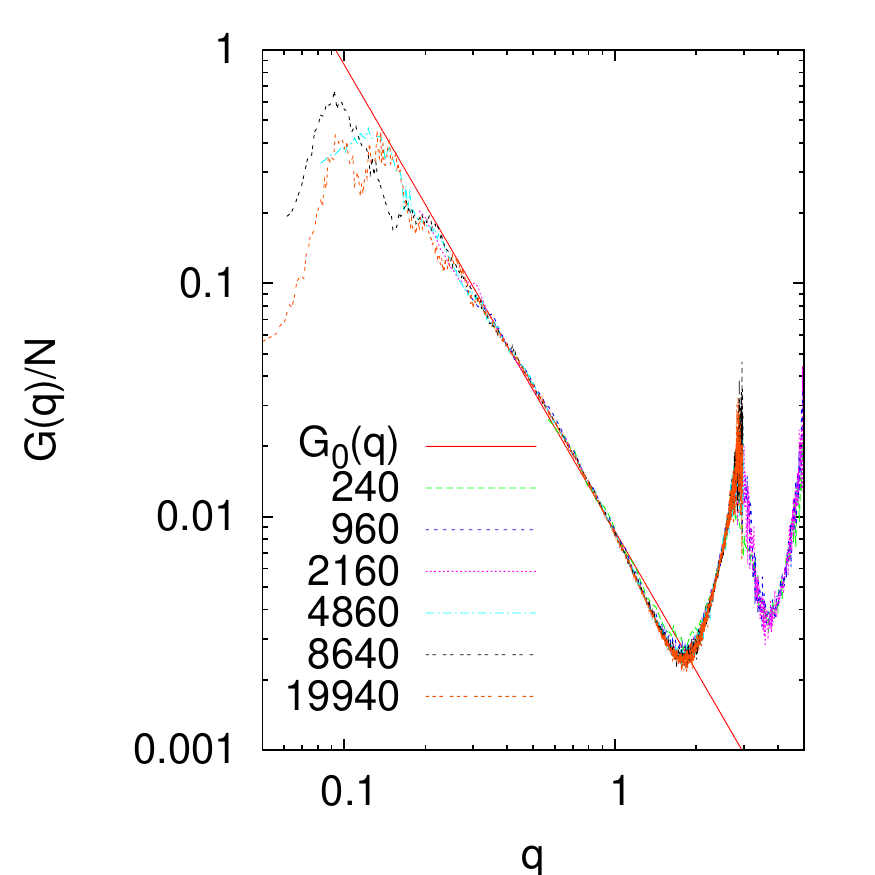}
\caption{Normal-normal correlation function $G(q)/N$ for all studied samples. The solid straight line gives the harmonic power law behaviour $G_0(q)/N$ with $\kappa$=1.1 eV. Deviations from harmonic behavior occur for $q$ close to the Bragg peaks at $q=4\pi/3a = 2.94$ \AA$^{-1}$  and $q=4\pi/\sqrt(3)a = 5.11$ \AA$^{-1}$  with $a=1.42$ \AA~ and at small $q$ where the peak of $G(q)$ at $q\approx 0.1$ signals a preferred length scale of about 70 \AA. }
\label{Gall}
\end{figure}

\begin{figure}
\includegraphics{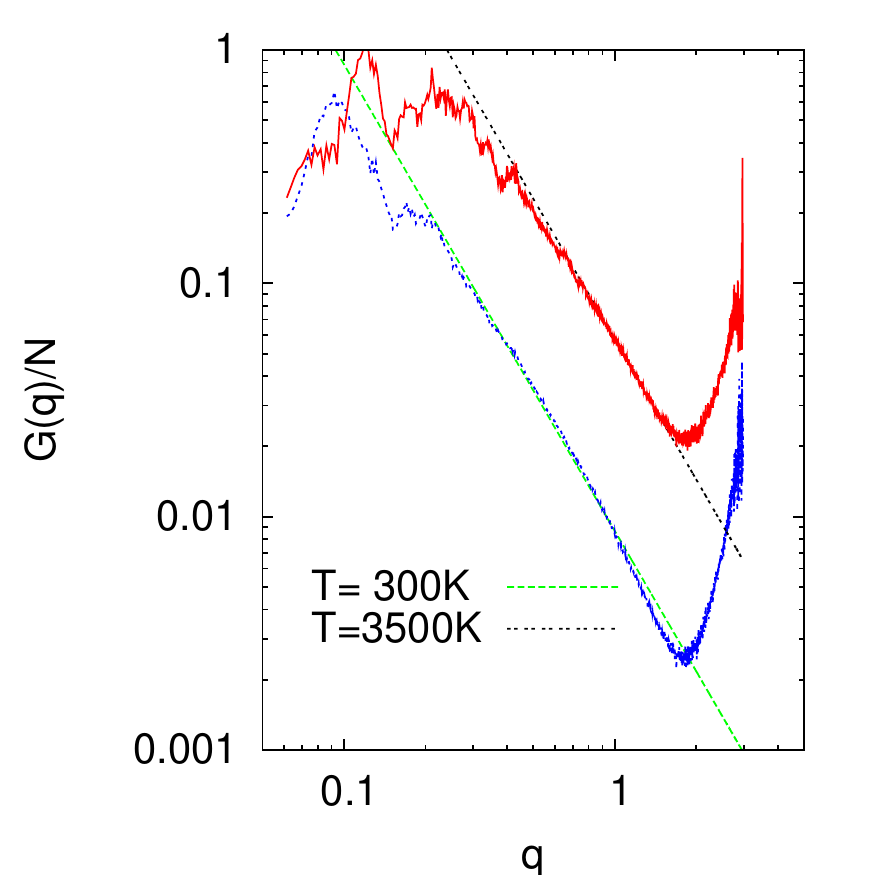}
\caption{Normal-normal correlation function $G(q)/N$ for the $N=8640$ sample at $T=300$ K and $T=3500$ K.
The solid straight line gives the harmonic power law behaviour $G_0(q)$ with $\kappa$=1.1 eV at $T=300$ and
$\kappa$=2.0 eV at $T=3500$ K.}
\label{G8640}
\end{figure}

\begin{figure}
\includegraphics[width=8cm]{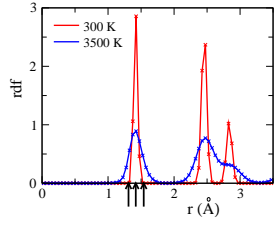}
\caption{Radial distribution function for the $N=8640$ sample at $T=300$ K and $T=3500$ K as a function of interatomic distances in ~\AA. The arrows indicate the length of double ($r=1.31$ \AA), conjugated  ($r=1.42$ \AA) and single ($r=1.54$ \AA) bonds.}
\label{rdf}
\end{figure}

In Fig.\ref{Gall} we show $G(q)$ at $T=300$ K for all considered samples
and in Fig.\ref{G8640} we compare the results at two temperatures for the
sample $N=8640$. First of all, there is a whole range of wave
vectors where the harmonic approximation given by Eq.(\ref{G0}) is quite
accurate. The interval is restricted above close to the Bragg
peaks and  in the limit of small $q$. The deviations are opposite
at the two sides. The rigidity $\kappa=1.1$  eV extracted from
the data at $T=300$ K by comparison with Eq.(\ref{G0}) is in very
good agreement with the experimental value of 1.2 eV derived from 
the phonon spectrum of graphite~\cite{nicklow}. Surprisingly, the same
comparison at $T=3500$ K gives a higher value $\kappa \approx 2.0$
eV whereas the contimuum theory~\cite{nelson} predicts the opposite trend
$\kappa(T)\approx\kappa-(3T/4\pi) \ln{(L/a)}$. This is due to the 
peculiar character of bonding in carbon. In the ground state of graphene all bonds 
are equivalent. However, even at room temperature, there is a large probability of 
having an asymmetric distribution of short/long (strong/weak) bonds, associated with 
local deviations from planarity. Indeed, the radial 
distribution functions shown in Fig.\ref{rdf} for both temperatures show a
broad distribution of first neighbors bond lengths, going down to the length of 
double bonds in carbon of 1.31 \AA~ at high temperature. Changes of bond conjugation are also the reason for 
the negative thermal expansion coefficient in graphene found 
in ab-initio calculations~\cite{thermal}. We observe a 0.13 \% contraction 
of the lattice spacing at $T=300$ K, in good agrement with the 0.11 \% of Ref.~\onlinecite{thermal}. 
We believe that it is the ability of carbon to form different types of bonding that
makes graphene different from a generic two-dimensional crystal. 

At small $q$, the behaviour of $G(q)$ is not described by the harmonic 
approximation $G_0(q)$ nor by the anharmonic expression $G_a(q)$.
The most remarkable feature of $G(q)$ is a maximum instead of the power law
dependence $G_a(q)$ that implies the absence of any
relevant length scale in the system.
The presence of this maximum,  instead, means that
there is a preferred average value of about 70~\AA. This length is also recognizable in real space images, as shown by the arrows in   Fig.\ref{snap}.  
Indeed, the two samples that are smaller than
this length do not show this decrease of $G(q)$ at low $q$. The
results at $T=3500$ K confirm this picture but with the maximum
shifted to a larger $q$ corresponding to a length of roughly $30$~\AA. 
This  temperature dependence of the typical ripple lengthscale
should be measurable.

The preferable length that we find  is reminiscent 
of the ``avoided criticality''
scenario in frustrated systems  near second-order phase transitions~\cite{tarjus}.
Such systems have an intrinsic tendency to be modulated and to
form some inhomogeneous patterns that destroys the scaling.
It is known that 
some soft condensed matter membranes tend
to spontaneous bending and ripple
formation~\cite{lubensky,katsnelson,manyuhina} but this 
behaviour for elemental solids like graphene is rather unexpected. 

The results obtained are relevant not only for a better
understanding of the stability and structure of graphene but also
of electronic transport. The fluctuations of normals leads to a
modulation of the hopping integrals and are bound to affect  the electronic
structure~\cite{morozov,castroneto}.  Knowledge of the
normal-normal correlation functions is necessary for the
calculation of the electron scattering by ripples.

The cleavage technique that has led to the discovery of graphene
has already been applied to other layered materials, like
BN~\cite{kostya0}, so that the investigation of structural
properties of one atom thick layers is important for a whole new
class of systems. We have found that even fluctuations at the scale of tens of interatomic distances cannot be described by continuum medium theory. 
 It will be very instructive to carry out
systematic experimental and theoretical investigations of other
two-dimensional crystals to understand  which properties are
common to flexible membranes and which ones are consequences of
particular features of the chemical bonding and interatomic
interactions.

{\it Acknowledgements}. We are thankful to Jan Kees Maan for suggestions and critical reading of the manuscript, Andre Geim, Kostya
Novoselov and Jannik Meyer for helpful discussions. This work was supported by the Stichting Fundamenteel Onderzoek der Materie (FOM) with
financial support from the Nederlandse Organisatie voor Wetenschappelijk Onderzoek (NW


\newpage
\vspace{0.3cm}
\begin{table}[ht]

\vspace*{0.2cm}
\begin{tabular}{|r|r|r|r|r|}
\hline
 $N$   &  p  &  q  & $L_x$(\AA) &   $L_y$(\AA)\\\hline
  240  &  10 &  6  &  24.59  &  25.56 \\
  960  &  20 & 12  &  49.29  &  51.12\\
 2160  &  30 & 18  &  73.78  &  76.68\\
 4860  &  45 & 27  & 110.68  & 115.02\\
 8640  &  60 & 36  & 147.57  & 153.36\\
19940  &  90 & 54  & 221.36  & 230.04\\
\hline
\end{tabular}
\caption{Details of the simulated samples. The initial, roughly squared, box is defined by $(L_x,L_y)=( 2p |${\bf a}${_1}|, q |${\bf a}${_1}+ 2 ${\bf a}${_2}|)$ where {\bf a}${_1}$ and {\bf a}${_2}$ are the in-plane lattice vectors 
{\bf a}${_1}= a \sqrt{3} ${\bf x}, {\bf a}${_2}= a \sqrt{3}/2 ${\bf x}$+ 3/2 ${\bf y}, 
{\bf x} and {\bf y} being cartesian unit vectors.}
\end{table}

\newpage

\newpage
\vspace*{0.5cm}

\newpage

\newpage

\begin{figure}
\includegraphics[width=8cm]{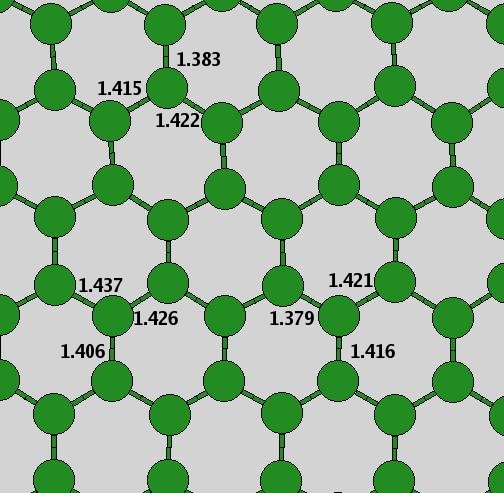}
\caption{Portion of one typical configuration of the $N=8640$ sample at $T=300$ K. The numbers indicate the bond length in \AA. Notice that often one of the bonds with first neighbours is much shorter than the other two. }
\label{zoom}
\end{figure}

\end{document}